\documentclass[preprint,longauthor]{aastex63}

\graphicspath{{./}{figures/}}

\begin{document}

\title{Detection of 49 Weak Dispersed Radio Pulses in a Parkes Observation of the X-ray Pulsar PSR J0537$-$6910}

\correspondingauthor{Fronefield Crawford}
\email{fcrawfor@fandm.edu}

\author[0000-0002-2578-0360]{Fronefield Crawford}
\affiliation{Department of Physics and Astronomy, Franklin and Marshall College, P.O. Box 3003, Lancaster, PA 17604, USA}

\begin{abstract}
I conducted a new search for dispersed radio pulses from the X-ray pulsar PSR J0537$-$6910 in the Large Magellanic Cloud in a long (11.6 hr) archival 1.4 GHz Parkes search observation. I searched dispersion measures (DMs) between 0 and 10000 pc cm$^{-3}$ and detected 49 pulses with a signal-to-noise ratio (S/N) greater than 7 at a wide range of DMs using the HEIMDALL and FETCH pulse detection and classification packages. All of the pulses were weak, with none having a S/N above 8.5. There was a significant excess of pulses observed in the DM range of the known pulsar population in the LMC, suggesting that these pulses may originate from LMC pulsars. Three repeat pulses, each having widths $\la 1$ ms, were detected in a single DM trial of 103.412 pc cm$^{-3}$, which is in the LMC DM range. This is unlikely to occur by chance in a single DM trial in this search at the (marginally significant) 4.3$\sigma$ level. It remains unclear whether any of the detected pulses in the sample are from PSR J0537$-$6910 itself. 
\end{abstract}

\keywords{Pulsars (1306); Radio Transient Sources (2008)}

\section{Introduction and Background} 

PSR J0537$-$6910 is a young rotation-powered X-ray pulsar associated with the supernova remnant (SNR) N157B in the Large Magellanic Cloud (LMC). It was first discovered as a 16-ms pulsed X-ray source by \cite{mgz+98}, and it is the most rapidly rotating unrecycled pulsar currently known. Despite previous searches for both periodic radio emission and single radio pulses with the Parkes 64-m radio telescope (``Murriyang'') \citep{ckm+98,cmj+05}, PSR J0537$-$6910 has not yet been detected as a radio emitter.

PSR J0537$-$6910 is a particularly good candidate to search for giant radio pulses. \citet{cst96} suggested a relationship between the giant pulse emission mechanism and the magnetic field strength at the light cylinder radius, defined as the equatorial radius at which co-rotation with the pulsar would equal the speed of light. This magnetic field strength can be computed from the spin parameters of the pulsar according to $B_{lc} = 3 \times 10^{8} \dot{P}^{1/2} P^{-5/2}$~G, where $P$ and $\dot{P}$ are the pulsar period in seconds and its time derivative, respectively. 

Along with two other young pulsars that are known emitters of giant radio pulses, the Crab pulsar \citep{sr68,lcu+95} and PSR B0540$-$69, which also resides in the LMC \citep{jr03}, PSR J0537$-$6910 has a large light-cylinder magnetic field strength. It is noteworthy that five millisecond pulsars (MSPs) with much different properties than these young pulsars but with large light-cylinder magnetic field values have also been observed to emit giant radio pulses \citep{cst96,rj01,jr03,jkl+04,kbm+05,kbm+06}. PSR J0537$-$6910 has a light-cylinder field strength that is more than twice as large as either of the next two highest light-cylinder-field pulsars (the Crab pulsar and the MSP PSR B1937+21), making it a good target to test this hypothesis.

\begin{figure}
%\centerline{\psfig{figure=b_lc_plot.eps}}
\begin{center}
\includegraphics[width=1.00\textwidth]{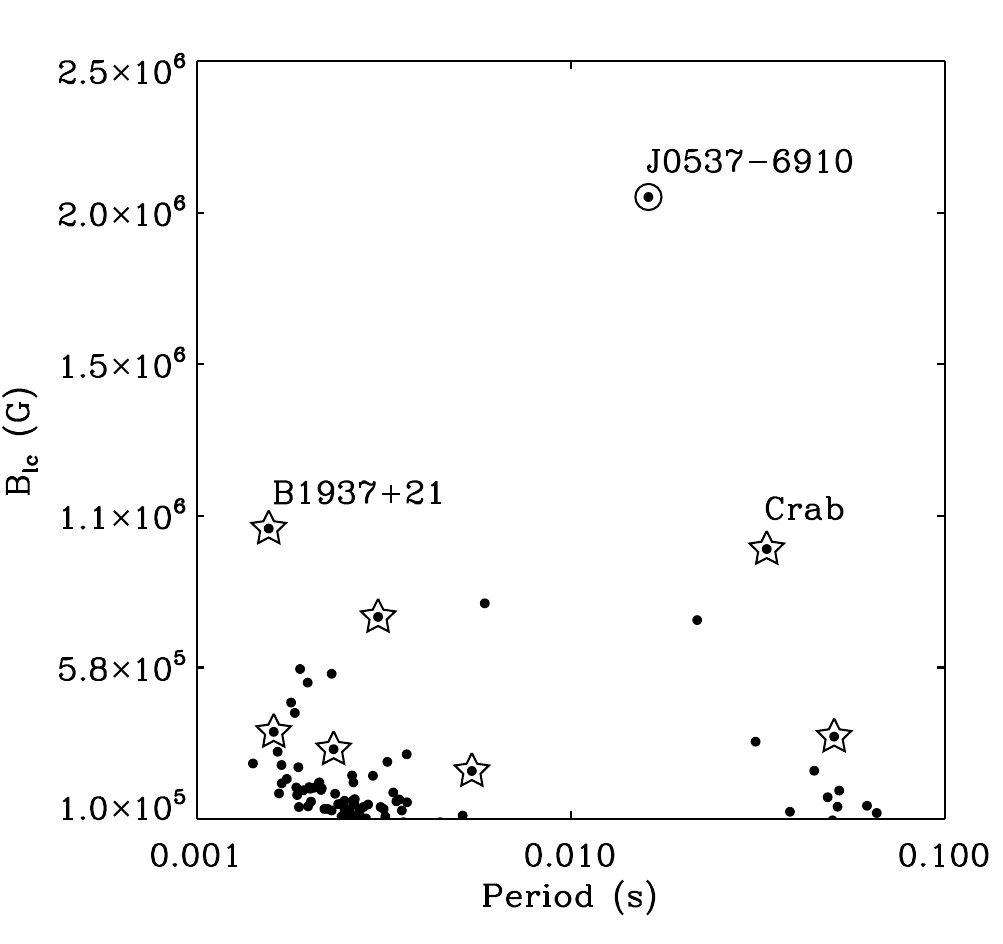}
\end{center}
\caption{Light-cylinder magnetic field strength vs.~spin period for pulsars in the ATNF pulsar catalog (version 1.70) \citep{mht+05} that have $B_{lc} > 10^{5}$ G. PSR J0537$-$6910 has the largest $B_{lc}$ by a factor of two and is indicated by the open circle. The seven pulsars observed to emit giant radio pulses are indicated by open stars; the two pulsars that have the next highest values of $B_{lc}$ after PSR J0537$-$6910 are labeled (PSR B1937+21 and the Crab pulsar). The population shown here can be divided into MSPs on the left and young pulsars on the right. See also Table \ref{tbl-1}.}
\label{fig-1}
\end{figure}

Table \ref{tbl-1} presents a current listing of the pulsars with the largest light-cylinder field strengths, with the observed giant radio pulse emitters identified. This is also illustrated in Fig. \ref{fig-1}, where pulsars with $B_{lc} > 10^{5}$ G are plotted as a function of the spin period (see also Table 1 of both \cite{cmj+05} and \cite{mc03}, and Fig.~4 of \cite{cst96}; however, these references do not have the more recently discovered pulsars shown in Table \ref{tbl-1} and Fig.~\ref{fig-1}). 

\section{Radio Search History}

Prior searches for both periodic and single-pulse radio emissions from PSR J0537$-$6910 were conducted (unsuccessfully) using Parkes at several frequencies. \citet{mc03} searched for the pulsar in a single 0.5 hr pointing at 435 MHz, while \citet{ckm+98} searched two 4 hr pointings at 660 MHz and a single 6-hr pointing at 1374 MHz. Subsequently, \citet{cmj+05} observed the pulsar with a longer 11.6 hr integration at a center frequency of 1390 MHz. This observation had a 256 MHz bandwidth split into 512 channels and was sampled at 80 $\mu$s. In all of these searches, no dispersion measures (DMs) above 300 pc cm$^{3}$ were searched. This encompassed the DMs of the known LMC pulsar population (which currently ranges from 45 to 273 pc cm$^{-3}$; see, e.g., \citealt{hcb+22}). No convincing astrophysical signals were seen in any of these prior searches. This last (and longest) observation is the one I have searched again with newer software packages over a wider range of DMs.  

\section{Data Analysis}

I reprocessed this single long Parkes observation with the HEIMDALL pulse detection package \citep{b12}\footnote{\url{https://sourceforge.net/projects/heimdall-astro}} at trial DMs ranging from 0 to 10000 pc cm$^{-3}$ in order to search for single-pulse events, including fast radio bursts (FRBs; \citealt{lbm+07}). A total of 1011 DM trials were produced and searched by HEIMDALL. Boxcar-matched filtering windows of $2^{n}$ samples, with $n$ ranging from 0 to 9, were applied to each dedispersed time series to maintain maximum sensitivity to pulses with widths up to $\sim 41$ ms. This is significantly larger than the widths expected for any pulses from PSR J0537$-$6910 given its 16 ms spin period. 

All of the pulses detected by HEIMDALL were then analyzed by FETCH.\footnote{\url{https://github.com/devanshkv/fetch}} FETCH is a pulse classifier that assigns a probability of being real to each detected pulse based on its morphology and characteristics \citep{aab+20}. FETCH rated each detected pulse using its Model A (see Table 4 of \citealt{aab+20}) and assigned a probability of being a real, astrophysical pulse of between 0 and 1. I also searched the data for periodicities at DMs ranging from 0 to to 5000 using PRESTO tools \citep{r01} in case other pulsars were present in the same beam. No promising periodicity signals were detected.

\section{Results and Discussion}

A total of 49 single pulses were detected with a signal-to-noise ratio (S/N) above 7 which had a probability assigned by FETCH that was greater than 0.5. All of these pulses were subsequently checked visually to ensure they were not obvious radio frequency interference (RFI) signals that had been misidentified by FETCH. All of the detected pulses were weak, with none having a S/N above 8.5. This corresponds to a fluence threshold of 0.6 Jy ms (for a putative pulse width of $W = 1$ ms; this sensitivity limit scales as $\sqrt{W}$). 
Table \ref{tbl-2} lists these 49 pulse detections with their characteristics.

\subsection{Possible False Detections}

As a check to see if these pulses might have been artificially generated by the software during the detection and classification process, I repeated the search over the same range of negative DMs (0 to $-10000$ pc cm$^{-3}$) with the same S/N threshold of 7 in order to see whether a significant number of spurious candidates would be produced. This exercise produced only one artificial negative-DM candidate that was classified as real (with S/N = 7.0). This is in contrast to the results of \cite{psv+22} and \citet{hcb+22} who each conducted a similar test on two different large-scale pulsar surveys and found more artificial, low-S/N candidates at negative DMs than corresponding positive-DM candidates. This indicated that their sample of detected low-S/N candidates might be largely artificial. 

\cite{nhs+23} reported several cases in which weak pulses detected from the repeating FRB 20200120E were not classified as real by FETCH. These pulses were ultimately determined to be real owing to their DM proximity to the FRB. In this case, real, weak pulses had been classified as RFI and missed (false negatives), but not the reverse (no false positives were reported). Thus, this same misclassification issue would not have produced spurious, false-positive detections in this sample of 49 pulses (although it could possibly have resulted in missing some real pulses). 

For comparison, in 36.6 hr of targeted observations of several SNRs that used a similar 1.4 GHz Parkes observing setup and the same analysis procedure used here, no pulses (spurious or otherwise) were detected and classified as real, apart from four pulses from a known bright pulsar in the vicinity \citep{c23}. 

\begin{figure}
%\centerline{\psfig{figure=dm_smear.eps,width=5in}}
\begin{center}
\includegraphics[width=1.00\textwidth]{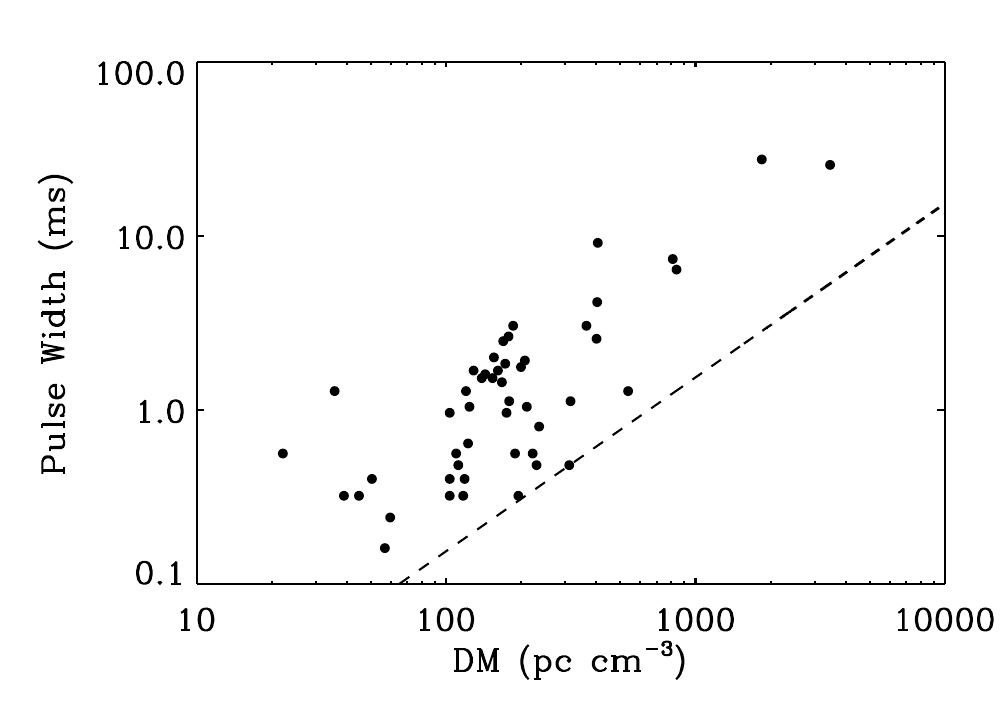}
\end{center}
\caption{Pulse width vs.~DM for 49 detected pulses. Also plotted is a dashed line indicating where the pulse width equals the dispersive smearing within the frequency channels. Real, astrophysical pulses would not be expected to lie below this line.}
\label{fig-4}
\end{figure}

Another indication that the pulses could be artificial is if their measured pulse widths are smaller than the corresponding dispersive smearing time within the finite frequency channels at that DM. Such narrow intrinsic pulses would be expected to be broadened by dispersion if the pulses were real. This dispersion smearing scales linearly with DM and is determined by $\tau = (202/f_{c})^{3}\, \Delta f\, {\rm DM}$, where $\tau$ is the smearing in ms, $f_{c}$ and $\Delta f$ are the center observing frequency and the channel width in MHz (1390 and 0.5, respectively), and DM is in pc cm$^{-3}$. Fig. \ref{fig-4} shows the measured pulse widths for the sample of 49 pulses from Table \ref{tbl-2} plotted as a function of DM, with the expected width from dispersive broadening also plotted. As seen in the figure, none of the detected pulses lie below the minimum (dispersive) width, lending further support to the notion that the pulses are not obviously artificial or otherwise generated by the software. Note that although scattering contributions to the broadening are not considered here, the Galactic contribution to scattering along this line of sight is negligible \citep{cl02}.

\subsection{Pulse Detection Rate} 

The average detection rate in this single observation was one pulse detected and classified as real above S/N of 7 for every 14 minutes of observing time. This rate is high compared to a similar search conducted of a large-scale survey of the LMC with Parkes using HEIMDALL and FETCH \citep{hcb+22}. That survey used a similar (though not identical) observing system that had a comparable raw sensitivity. They searched 702 beams totaling 1677 hr for single pulses out to a DM of 10000 pc cm$^{-3}$, and they used a similar maximum boxcar width (33 ms). A total of 229 pulses were found in that survey with a S/N above 7, a DM above 50 pc cm$^{-3}$, and a FETCH probability of being real above 0.9 (this included nine pulses detected from the giant pulse emitter PSR B0540$-$69). When using these same cutoffs and filtering criteria for the set of detections reported here, 33 of the 49 original pulses are retained. However, these were detected over the much smaller 11.6 hr of integration time. The pulse detection rate per unit of observing time is therefore $\sim 20$ times larger than the large-scale LMC survey analysis of \citet{hcb+22}. Some of this difference may be attributable to the fact that the lone observation analyzed here targeted the 30 Doradus star formation region, where more pulsars may be present than on average in the LMC. However, this large difference remains difficult to reconcile completely if the pulses detected here are indeed real and coming from the LMC. This discrepancy becomes even more pronounced if some fraction of the weak pulses detected by \citet{hcb+22} are not actually real (see the discussion above).

As outlined in \citet{cmj+05}, if the Crab pulsar were located at the distance of the LMC, it would emit a giant pulse every 20 minutes that would be detectable with this observing system. This would lead to several dozen such detections in this single observation. \citet{cmj+05} also indicate that a giant pulse from PSR B0540$-$69 in the LMC should be detectable every 0.5 hr with such an observing setup. This is broadly consistent with the detection of nine pulses from PSR B0540$-$69 (four of which were above S/N of 9) in a single 2.4-hr Parkes survey beam of the LMC which covered the location of that pulsar (\citealt{hcb+22}; see their Table 1). The sample reported here clearly does not include any pulses from a single source (such as PSR J0537$-$6910) with this brightness or frequency of occurrence.

\subsection{DM Distribution of Detected Pulses}

\begin{figure}
%\centerline{\psfig{figure=snr_vs_dm_idl_snr7.eps,width=5in}}
\begin{center}
\includegraphics[width=1.00\textwidth]{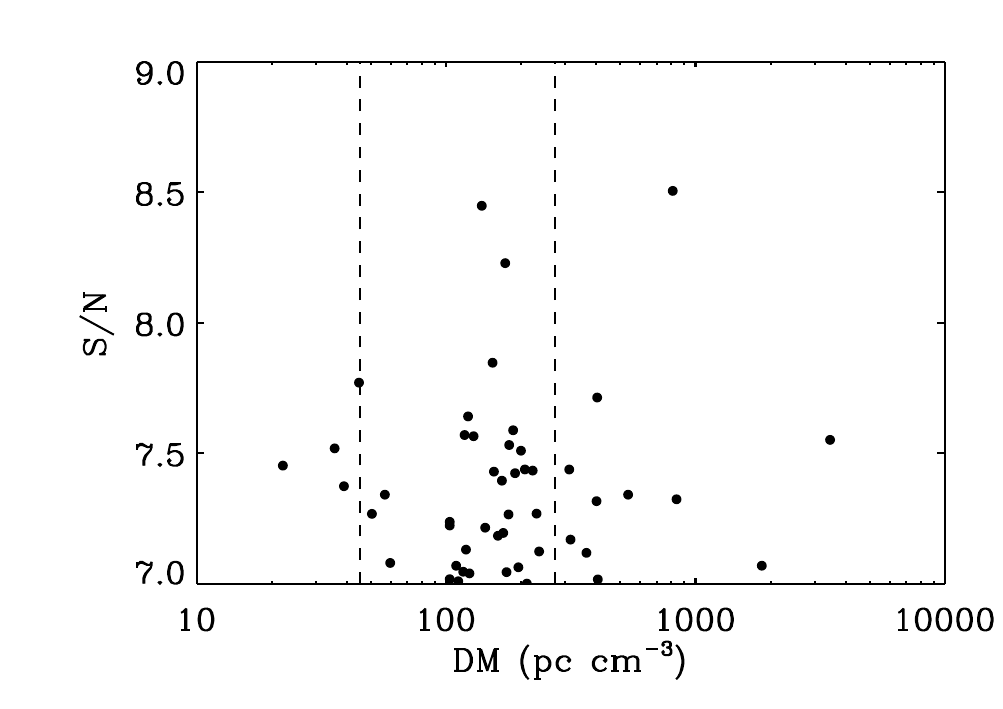}
\end{center}
\caption{Pulse S/N vs.~DM for 49 single-pulse detections with S/N above 7 which had a FETCH-assigned probability of being real greater than 0.5. There is an excess of pulses in the observed DM range of the known LMC pulsar population (45 to 273 pc cm$^{-3}$, indicated by the dashed vertical lines), suggesting that some of these pulses could be from LMC sources. Whether PSR J0537$-$6910 is the source of any of these pulses remains speculative.}
\label{fig-2}
\end{figure}

The largest DM of any currently known radio pulsar in the LMC is 273 pc cm$^{-3}$ \citep{rcl+13}, but the remainder of the known LMC population has DMs that lie between 45 and 147 pc cm$^{-3}$ \citep{mfl+06,rcl+13}. Fig. \ref{fig-2} shows our 49 detected pulses with S/N plotted against DM. The majority of the detected pulses (34 out of 49) fall within the observed DM range of the currently known LMC pulsar population. However, only 26\% of the DM trials in the search were in this range, and so we would expect only 13 events to occur here by chance. Given this, the likelihood of detecting 34 or more pulses by chance in the DM trials in this range is less than $10^{-6}$, 
%(about 5$\sigma$ significance)
suggesting that the observed excess is real. Therefore, it is possible that many of these pulses are from as-yet-unidentified pulsars in the LMC.

The wide distribution of DMs of the detected pulses indicates that they cannot all be coming from the same object. Some may be coming from other, as-yet-unidentified LMC pulsars (possible, given the excess of pulses seen in the LMC DM range). The few pulses with very large DMs (4 of the 49 had DM $> 800$ pc cm$^{-3}$; see Table \ref{tbl-2} and Fig. \ref{fig-2}) could be FRBs originating from well beyond the LMC.

\subsection{Repeat Pulses}

\begin{figure}
%\centering
\begin{center}
\includegraphics[width=0.30\textwidth]
{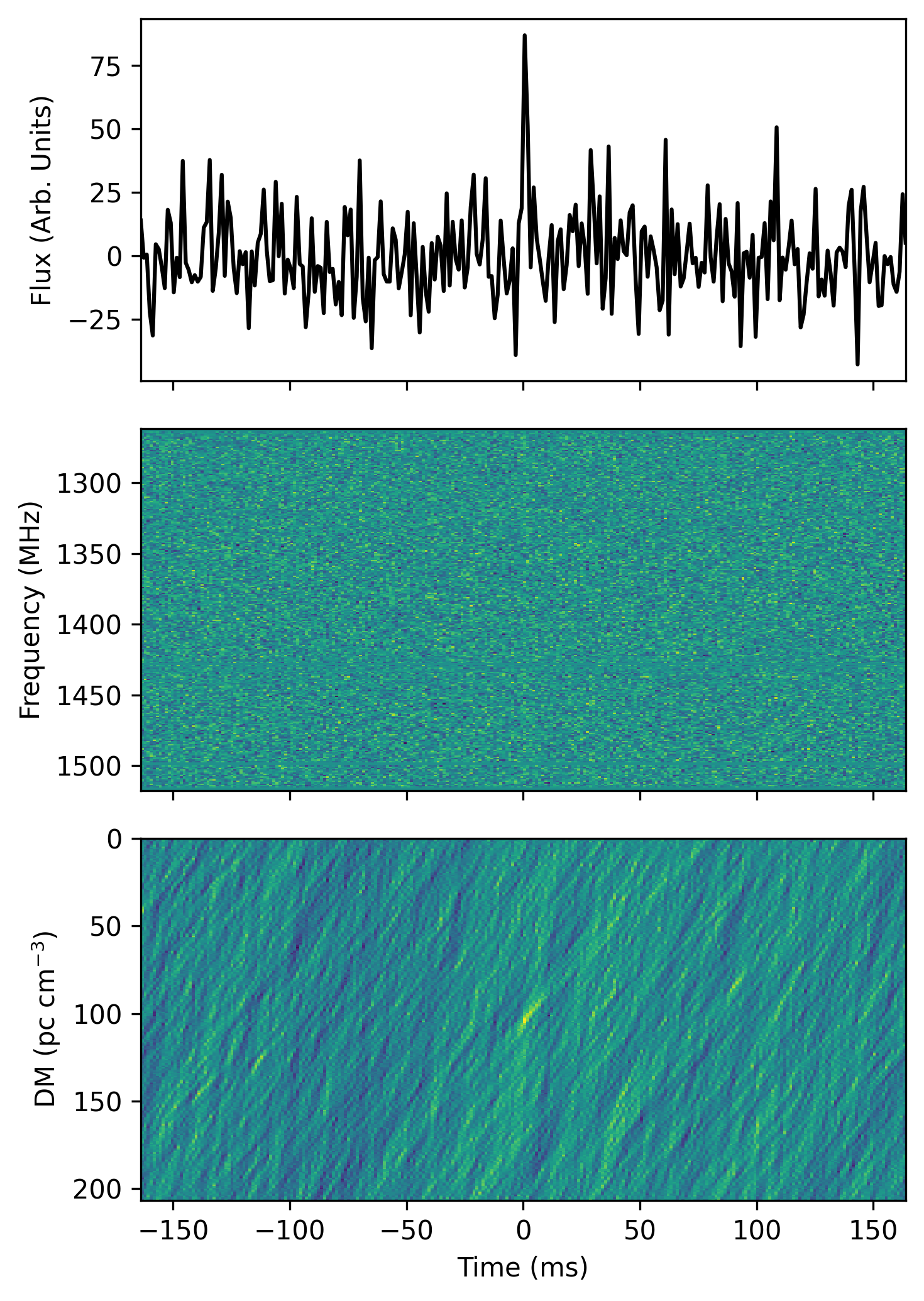}
\includegraphics[width=0.30\textwidth]
{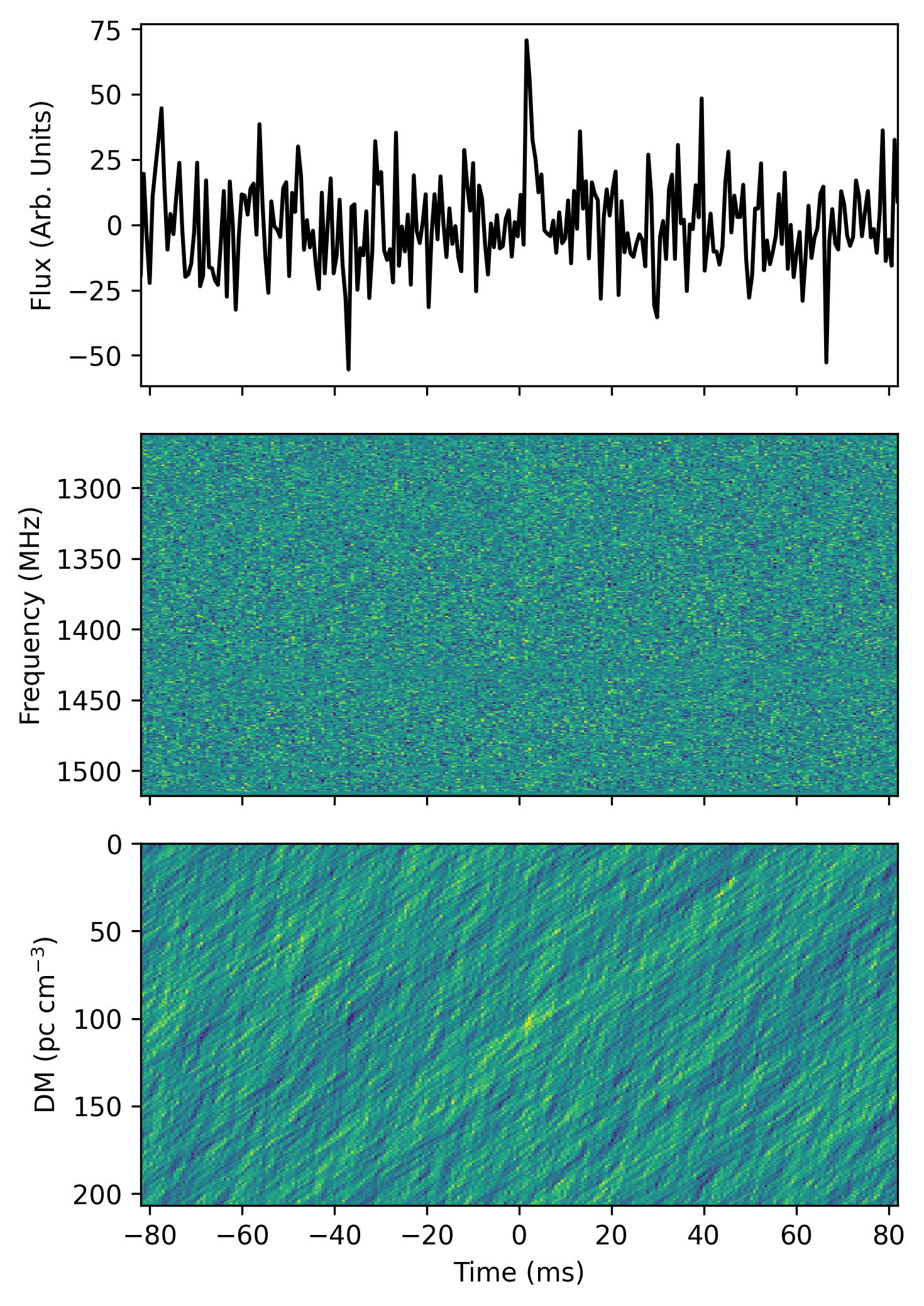}
\includegraphics[width=0.30\textwidth]{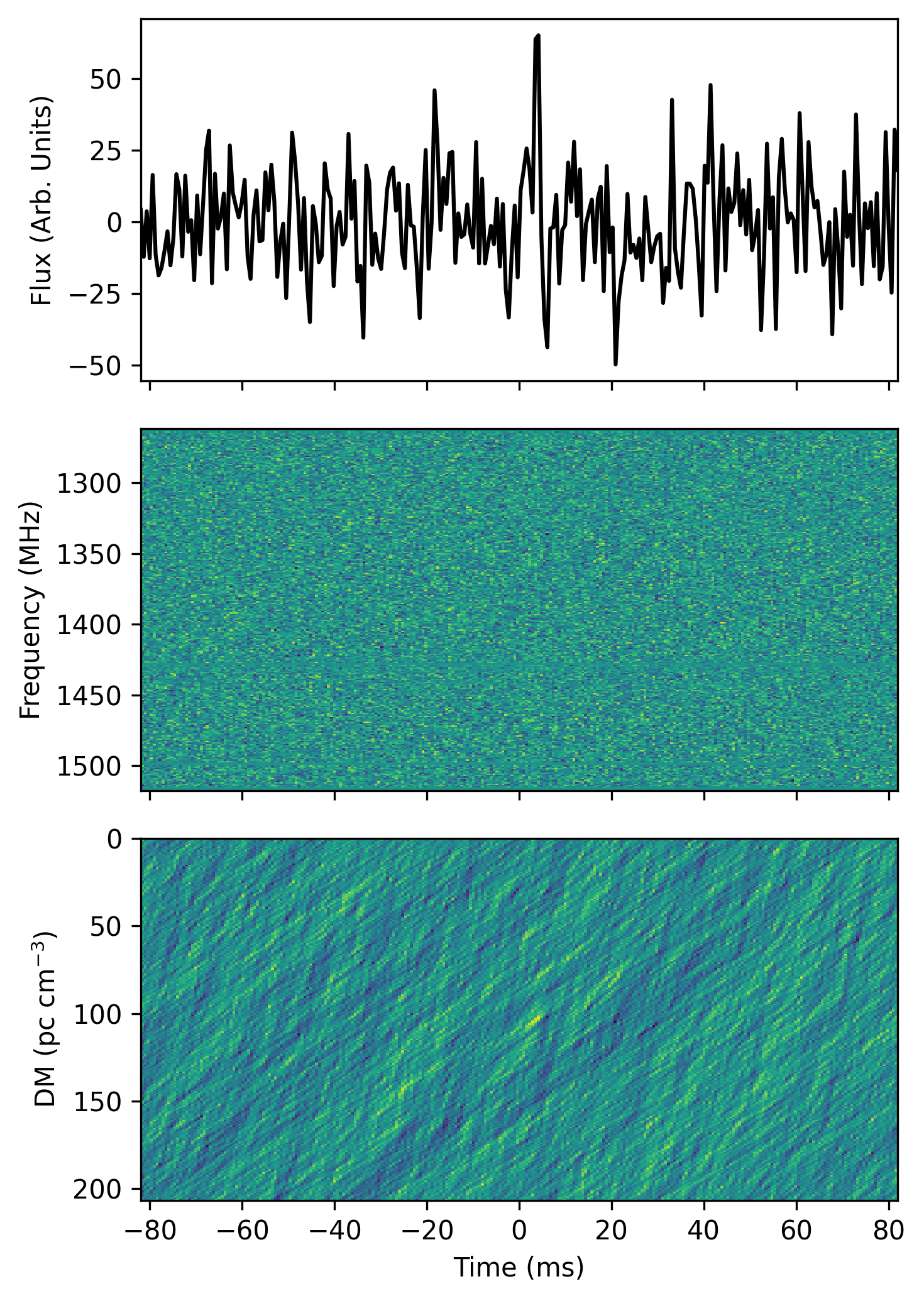}
\end{center}
\caption{Three weak single pulse detections occurring in a single DM trial of 103.412 pc cm$^{-3}$. The pulses are shown in the order given in Table \ref{tbl-2}, which provides further details. The top panel in each case shows flux vs.~time, with the pulse centered at time zero. The middle panel shows frequency vs.~time  after dedispersion has been applied. A broadband signal (straight vertical line) would be expected in this panel for a typical pulse from a pulsar. The bottom panel shows  DM vs.~time. A localized signal appearing at a non-zero DM would be expected for an astrophysical signal. The likelihood of three or more of the 49 detected pulses appearing in a single DM trial by chance in this search is small (unlikely at the 4.3$\sigma$ level). All three pulses have comparable widths and could be coming from the same pulsar in the LMC, though probably not from PSR J0537$-$6910 (see the discussion in the main text).}
\label{fig-3}
\end{figure}

Three pulses were detected in a single DM trial (DM = 103.412 pc cm$^{-3}$). These three pulses are shown in Fig. \ref{fig-3} and are identified in Table \ref{tbl-2}. The likelihood of three or more pulses occurring by chance in a single DM trial can be estimated using the total number of pulses detected (49) and the total number of DM trials in the search (1011). Following the analysis outlined in Section 4.2 of \citet{phl+24} for a similar likelihood estimate for a single-pulse search of M82, I determined that this is unlikely to occur by chance at the (marginally significant) 4.3$\sigma$ level. No other DM trial had any repeat pulses (see Table \ref{tbl-2}). 

This DM value is well within the observed DM range for LMC pulsars, and all three pulses had comparable widths of between 0.3 and 1.0 ms, as determined from the HEIMDALL detections (see Table \ref{tbl-2}). Note that the dispersive smearing within frequency channels at this DM is 0.15 ms. This is significantly smaller than the measured pulse widths (see also Fig. \ref{fig-4}), indicating that these measured widths are largely intrinsic to the pulsar. This also suggests that these are not micropulses or nanoshots like those seen in giant pulses from the Crab pulsar \citep{he07,heg16}. 

It is possible that these three pulses could all be coming from the same pulsar in the LMC. The pulse widths are much smaller than the 16 ms period of PSR J0537$-$6910, so they could be coming from PSR J0537$-$6910 specifically. However, the time separations between the three pulses are not close to an integer number of pulse periods from PSR J0537$-$6910. To determine this, the topocentric period and its uncertainty were determined for PSR J0537$-$6910 from the ATNF catalog parameters \citep{mht+05}\footnote{\url{https://www.atnf.csiro.au/research/pulsar/psrcat/}}
using PRESTO tools. The topocentric period uncertainty was combined with the measured half-widths of the pulses to obtain an uncertainty in the time of each pulse separation. In all three cases, the uncertainty (which was 2\%-3\% of the pulse period of PSR J0537$-$6910) was much less than the remainder when the pulse time difference was divided by the pulsar period (these remainders were 41\%, 18\%, and 77\%). This disfavors the conclusion that the pulses are from PSR J0537$-$6910. However, given the large and prolific timing glitches seen for the pulsar (e.g., \citealt{hea+20}), this may not be definitive. I also dedispersed the raw data at this DM and folded the resulting time series using the ephemeris of PSR J0537$-$6910. No signal was detected in this fold.

\section{Conclusions}

In a new analysis of an archival Parkes search observation of the LMC X-ray pulsar PSR J0537$-$6910, I detected 49 dispersed single radio pulses with a S/N greater than 7. All 49 pulses had a FETCH likelihood of being real that was greater than 0.5. None of the 49 detected pulses had a S/N above 8.5, corresponding to a fluence threshold of 0.6 Jy ms (for a putative 1 ms pulse width). A significant excess of the detected pulses (34 out of 49) occurred within the DM range of the known LMC pulsar population, suggesting that some of these pulses may be from as-yet-unidentified LMC pulsars or possibly from PSR J0537$-$6910 itself. Three pulses having widths $\la 1$ ms were detected in a single DM trial (DM = 103.412 pc cm$^{-3}$). This is unlikely to occur by chance at the 4.3$\sigma$ level. This DM value is within the observed range for LMC pulsars, suggesting that the pulses may originate from a pulsar in the LMC. Future observations with more sensitive, next-generation facilities may be useful for determining whether any of the pulses detected in the sample are from PSR J0537$-$6910.

\acknowledgments
The Parkes radio telescope is part of the Australia Telescope National Facility (grid.421683.a), which is funded by the Australian Government for operation as a National Facility managed by CSIRO. This work was supported in part by National Science Foundation (NSF) Physics Frontiers Center award Nos. 1430284 and 2020265, and used the Franklin and Marshall College compute cluster, which was funded through NSF grant 1925192. 

\vspace{5mm}
\facilities{Parkes}

\software{HEIMDALL \citep{b12,bbb+12},  FETCH 
\citep{aab+20}
          }

\bibliography{c24}{}
\bibliographystyle{aasjournal}

%TABLES

\begin{deluxetable}{lccc}
\tablecaption{Cataloged Pulsars with the Largest Light-Cylinder Magnetic Field Strengths}
\tablewidth{0pt}
\tablehead{
\colhead{PSR} \vspace{-0.1cm} &  \colhead{$B_{lc}$} & \colhead{Type} & \colhead{Giant Pulse} \\
\colhead{} & \colhead{($10^{5}$ G)} & \colhead{} & \colhead{References} 
}
\startdata
J0537$-$6910          & 20.7        & young$^{a}$       &               \\
{\bf B1937+21}        & {\bf 10.2}  & {\bf MSP}   & \citet{cst96} \\ 
{\bf B0531+21 (Crab)} & {\bf 9.6}   & {\bf young} & \citet{sr68}  \\ 
J1402+13              & 7.8         & MSP$^{a}$        &               \\
{\bf B1821$-$24A}     & {\bf 7.4}   & {\bf MSP}   & \citet{rj01}  \\
J0058$-$7218          & 7.3         & young$^{a}$       &               \\
J1748$-$2446ak        & 5.8         & MSP         &               \\
J1701$-$3006F         & 5.6         & MSP         &               \\
J1737$-$0314A         & 5.3         & MSP         &               \\
J1555$-$2908          & 4.7         & MSP         &               \\
J1835$-$3259B         & 4.4         & MSP         &               \\
{\bf B1957+20}        & {\bf 3.8}   & {\bf MSP}   & \citet{jkl+04} \\
{\bf B0540$-$69}      & {\bf 3.6}   & {\bf young} & \citet{jr03}  \\
J1400$-$6325          & 3.5         & young       &               \\
{\bf J0218+4232}      & {\bf 3.2}   & {\bf MSP}   & \citet{kbm+06} \\ 
\enddata 
\tablecomments{$^{a}$Not a radio pulsar. Data were taken from the ATNF Pulsar Catalog (version 1.70). The top fifteen pulsars ranked by $B_{lc}$ are listed. Bold entries indicate pulsars observed to emit giant radio pulses. Note that one other pulsar, PSR B1820$-$30A, is an MSP that emits observable giant radio pulses \citep{kbm+05}, but it is not listed here since it ranks 23rd on this list. See also Fig.~\ref{fig-1}.} 
\label{tbl-1}
\end{deluxetable}

\begin{deluxetable}{ccccrc}
\tabletypesize{\scriptsize}
\tablecaption{List of Detected Single Pulses Ordered by DM}
\tablewidth{0pt}
\tablehead{
\colhead{Pulse} \vspace{-0.1cm} &  \colhead{Time of}  & \colhead{DM Trial} & \colhead{S/N} & \colhead{FETCH} & \colhead{Pulse Width} \\
\colhead{Number}  & \colhead{Pulse (s)} & \colhead{(pc cm$^{-3}$)} & \colhead{} & \colhead{Likelihood} & \colhead{(ms)}
}
\startdata
%1 & 5924.5415(5)  & 103.412 & 7.2 & 99.97\%  & 1.0 \\ %0.99969757 
%2 & 6764.7114(2)  & 103.412 & 7.2 & 98.10\% & 0.3 \\ %0.98097205  
%3 & 40538.7034(2) & 103.412 & 7.0 & 62.63\% & 0.5 \\ %0.62849563 
1  & 6676.2417(3)    & 22.161  & 7.5 & 99.997\% & 0.6 \\
2  & 2442.0319(6)    & 35.742  & 7.5 & 99.985\% & 1.3 \\
3  & 8397.3183(2)    & 38.948  & 7.4 & 99.841\% & 0.3 \\
4  & 5369.5697(2)    & 44.762  & 7.8 & 99.515\% & 0.3 \\
5  & 20223.2706(2)   & 50.421  & 7.3 & 71.246\% & 0.4 \\
6  & 40026.1995(1)   & 56.835  & 7.3 & 99.988\% & 0.2 \\
7  & 3497.6510(1)    & 59.695  & 7.1 & 99.438\% & 0.2 \\
8* & 5924.5415(5)    & 103.412 & 7.2 & 99.970\% & 1.0 \\
9* & 6764.7114(2)    & 103.412 & 7.2 & 98.097\% & 0.3 \\
10*& 40538.7033(2)   & 103.412 & 7.0 & 62.850\% & 0.4 \\
11 & 26744.4141(3)   & 109.731 & 7.1 & 99.983\% & 0.6 \\
12 & 5523.0110(2)    & 111.903 & 7.0 & 99.886\% & 0.5 \\
13 & 2083.5429(2)    & 117.103 & 7.0 & 99.990\% & 0.3 \\
14 & 24753.1893(2)   & 118.623 & 7.6 & 52.817\% & 0.4 \\
15 & 11974.3024(6)   & 120.160 & 7.1 & 99.945\% & 1.3 \\
16 & 8762.3732(3)    & 122.494 & 7.6 & 99.989\% & 0.6 \\
17 & 18975.3036(5)   & 124.071 & 7.0 & 77.577\% & 1.0 \\
18 & 22657.4292(8)   & 128.902 & 7.6 & 66.141\% & 1.7 \\
19 & 5913.4912(8)    & 139.035 & 8.4 & 99.758\% & 1.5 \\
20 & 8012.5940(8)    & 143.452 & 7.2 & 99.981\% & 1.6 \\
21 & 27263.3490(8)   & 153.597 & 7.8 & 96.984\% & 1.5 \\
22 & 36148.5644(10)  & 155.507 & 7.4 & 99.468\% & 2.0 \\
23 & 1460.5734(8)    & 161.362 & 7.2 & 99.995\% & 1.7 \\
24 & 12319.5048(7)   & 167.412 & 7.4 & 99.998\% & 1.4 \\
25 & 26833.7275(12)  & 169.473 & 7.2 & 89.701\% & 2.5 \\
26 & 18547.8670(9)   & 172.608 & 8.2 & 98.323\% & 1.8 \\
27 & 9213.3246(5)    & 174.726 & 7.0 & 96.529\% & 1.0 \\
28 & 14323.1635(13)  & 177.947 & 7.3 & 59.221\% & 2.6 \\
29 & 15232.4035(6)   & 179.033 & 7.5 & 94.064\% & 1.1 \\
30 & 26228.9132(15)  & 185.673 & 7.6 & 99.998\% & 3.0 \\
31 & 40129.3022(3)   & 189.077 & 7.4 & 90.696\% & 0.6 \\
32 & 34427.6883(2)   & 194.876 & 7.1 & 79.459\% & 0.3 \\
33 & 38263.8594(9)   & 199.633 & 7.5 & 93.066\% & 1.8 \\
34 & 21103.5138(10)  & 206.969 & 7.4 & 99.992\% & 1.9 \\
35 & 6778.7168(5)    & 210.730 & 7.0 & 61.309\% & 1.0 \\
36 & 20002.8843(3)   & 222.397 & 7.4 & 88.357\% & 0.6 \\
37 & 1891.0116(2)    & 230.508 & 7.3 & 97.514\% & 0.5 \\
38 & 22474.6054(4)   & 236.069 & 7.1 & 99.995\% & 0.8 \\
39 & 32339.2015(2)   & 311.730 & 7.4 & 85.183\% & 0.5 \\
40 & 5607.8609(6)    & 315.418 & 7.2 & 94.921\% & 1.1 \\
41 & 12581.6178(15)  & 365.243 & 7.1 & 99.765\% & 3.0 \\
42 & 7558.9811(13)   & 401.071 & 7.3 & 94.033\% & 2.6 \\
43 & 39081.7106(21)  & 403.421 & 7.7 & 99.410\% & 4.2 \\
44 & 12254.4407(46)  & 405.784 & 7.0 & 52.500\% & 9.1 \\
45 & 33980.0101(6)   & 536.747 & 7.3 & 99.978\% & 1.3 \\
46 & 18378.7853(37)  & 810.519 & 8.5 & 99.847\% & 7.4 \\
47 & 16925.0990(32)  & 839.202 & 7.3 & 99.502\% & 6.4 \\
48 & 21693.3493(138) & 1844.050& 7.1 & 91.929\% & 27.5\\
49 & 26754.9458(128) & 3463.760& 7.6 & 99.987\% & 25.6\\
\enddata 
\tablecomments{The time of the pulse is relative to the start of the integration at MJD 52888.61267361. The figure in parentheses represents the uncertainty in the last digit of the time of the pulse as determined by HEIMDALL. The pulse widths were also measured from the HEIMDALL detections. Pulses 8, 9, and 10 (indicated with asterisks) are the three pulses that appeared in a single DM trial. See also Fig. \ref{fig-3}.}
\label{tbl-2}
\end{deluxetable}

%FIGURES

\end{document}